\documentclass[letterpaper]{appolb}
\usepackage{graphicx}
\newcommand{\be}{\begin{equation}}
\newcommand{\ee}{\end{equation}}
\newcommand{\mom}{\frac{d^d k}{(2\pi)^d}}
\newcommand{\spc}{d^d x}
\begin{document}
\pagestyle{plain}
\eqsec
\newcount\eLiNe\eLiNe=\inputlineno\advance\eLiNe by -1
\title{Large N phase transitions under scaling and their uses
\thanks{Presented at {\sl Cracow School of Theoretical Physics}, Zakopane, 5.31 -- 6.10, 2009}}
\author{H.~Neuberger
\address{Department of Physics and Astronomy,\\ Rutgers University,  Piscataway, NJ 08854-0849, U.S.A} }
\maketitle 

\begin{abstract}The eigenvalues of Wilson loop matrices in $SU(N)$ gauge
theories in dimensions 2,3,4 at infinite $N$ 
are supported on a small arc on the unit
circle centered at $z=1$ for small loops, but expand to the entire unit circle 
for large loops. These two regimes are separated by a large $N$ 
phase transition whose universal properties are the same in $d=2,3$ and $4$. 
Hopefully, this large $N$ universality 
could be exploited to bridge traditional
perturbation theory calculations, valid for small loops, with effective string
calculations for large loops. A concrete case of such a calculation would obtain
analytically an estimate of 
the large $N$ string tension in terms of the perturbative
scale $\Lambda_{SU(N)}$.
\end{abstract}

\section{Introduction}

These lecture notes provide an elementary introduction to the topic described
in the abstract. Most analytic results are presented as exercises -- the
solutions of most of 
these exercises can be found in the papers in the reference list.
These derivations are not reproduced in the notes. Rather than describing
in detail the numerical results, the focus is on the general logic of the 
numerical tests. Again, details are in the references and actual numbers
and graphs are not reproduced in the notes. 
These notes reflect my personal viewpoint. 

The entire research topic rests on a 
paper by Durhuus and Olesen,~\cite{DO}, 
from 1981, who found a large $N$ phase transition in two dimensional YM. 
Much of the work on the crossover was influenced by Blaizot and Nowak who
pointed out almost two years ago 
that one might view the onset of confinement as an analogue of 
the onset of Burgers' turbulence~\cite{blaizot}.

\section{Abelian Wilson loop operators}

We work in the framework of Euclidean Field Theory, in $R^d$, where d=2,3,4. 
The focus is on pure gauge theory, that is, there is no matter. 

\subsection{Free abelian gauge theory}

We start with $U(1)$ gauge theory,
a free, Gaussian, theory. The quantity below, $Z[J_\mu ]$, 
contains all of the information about this theory:
\be
Z[J_\mu]=\int \left [ dA_\mu\right ] e^{-\frac{1}{4g^2}\int\spc F_{\mu\nu}^2 + i \int\spc J_\mu A_\mu}\equiv \int \left [ dA_\mu\right ] e^{-S[A_\mu] + i \int\spc J_\mu A_\mu}
\ee
All vector indices $\mu(\nu)$ are pairwise contracted.
The field strength $F_{\mu\nu}$ is defined by
\be
F_{\mu\nu}=\partial_\mu A_\nu - \partial_\nu A_\mu ~~{\rm and}~~F^2_{\mu\nu} = F_{\mu\nu} F_{\mu\nu}
\ee

\subsection{Carrying out the integral over $A_\mu$}
\subsubsection{Decoupling the modes}
The integral is Gaussian and there is translational invariance, so we can decouple the
modes by Fourier transforming the fields.
\be
A_\mu (x) =\int\mom e^{-ikx} {\tilde A}_\mu (k);~~~{\tilde A}_\mu (k) = {\tilde A}_\mu^\ast (-k)
\ee
\be
Z[J_\mu] = \int \prod_k \left [d{\tilde A}_\mu (k) \right ] 
e^{-\frac{1}{2g^2}\int \mom {\tilde A}_\mu (k ) [ \delta_{\mu\nu} k^2 - k_\mu k_\nu ] {\tilde
A}_\nu (-k) } e^{i\int\mom {\tilde J}_\mu (-k) {\tilde A}_\mu (k) }
\ee

We now decompose the vectors ${\tilde A}_\mu (k)$
\be
{\tilde A}_\mu (k) =\frac{k_\mu}{|k|} a_L (k)+\epsilon_\mu^i (k) a_\perp^i (k)
\ee
with 
\be
\epsilon_\mu^i k_\mu=0,~~~\epsilon_\mu^i (k) \epsilon_\mu^j (k)=\delta^{ij},~i,j=1,..,d-1
\ee
The quadratic term in the exponent of the path integral does not depend on $a_L(k)$. 
Hence, if the current obeys $k_\mu{\tilde J}_\mu (k)=0$ or, equivalently
$\partial_\mu J_\mu (x)=0$, one can introduce another weight factor 
for $a_L(k)$ to make the integral per $k$-mode finite, without $Z[J_\mu ]$ depending
on the details of that weight factor. One can even drop the $a_L(k)$ variables altogether.

\subsubsection{Current conservation}

If one introduces a $J_\mu(x)$ which is not conserved, that is 
$\partial_\mu J_\mu (x)\ne 0$, the integrand does not have a local extremum around which
to expand. The first variation of the exponent gives
\be
\partial_\mu F_{\mu\nu} = g^2 J_\nu 
\ee
If the above
equation for $A_\mu(x)$ had a solution, this would imply
that $\partial_\mu J_\mu (x)=0$ by virtue of $F_{\mu\nu}=-F_{\nu\mu}$.

One could interpret the $a_L$ integration as forcing the longitudinal component
of ${\tilde J_\mu}$ to vanish. The theory is simply restricted to probing by conserved 
external currents only.

More generally, only gauge invariant observables are meaningfully defined by the
theory. The gauge transformation is
\be
{\tilde A}_\mu (k)\rightarrow {\tilde A}_\mu (k) -i k_\mu \chi(k)
\ee
It only affects $a_L$; gauge invariant observables do not depend on $a_L$. 

The most localized conserved current one could imagine is
\be
J_\mu (x) = \int_0^l d\tau \delta^d (x-z(\tau))\frac{dz_\mu}{d\tau}
\ee
where $z_\mu(\tau)$ is a closed contour described in $R^d$  by $\tau$
varying from $\tau=0$ to $\tau=l$. The current is defined by the curve
itself, not by its parameterization. We fix the latter by
\be
\left (\frac{dz_\mu}{d\tau}\right)^2=1
\ee
This makes $l$ the perimeter of the curve, since
\be
\int_0^l d\tau \sqrt{
\left (\frac{dz_\mu}{d\tau}\right)^2}=l
\ee
The curve had to be closed to ensure current conservation:
\begin{eqnarray}
&\partial_\mu J_\mu (x) =\int_0^l d\tau \frac{dz_\mu}{d\tau} \frac{\partial}{\partial x_\mu}
\delta^d (x-z(\tau))=-\int_0^l d\tau \frac{d}{d\tau}\delta^d(x-z(\tau))=\\&
\delta^d(x-z(0))-\delta^d(x-z(l))=0
\end{eqnarray}

In Fourier space
\be
{\tilde J}_\mu (k) = \int \spc e^{i kx} J_\mu (x) =
\int_0^l d\tau\frac{dz_\mu}{d\tau} e^{ikz(\tau)}
\ee
Current conservation now is seen from
\be
k_\mu {\tilde J}_\mu (k) = -i \int_0^l d\tau \frac{d}{d\tau} e^{ikz(\tau)}=0
\ee

\subsubsection{Wilson loop operator}

The Wilson loop operator is $W[A]=e^{i\int \spc J_\mu A_\mu}=e^{i\oint dz_\mu A_\mu(z)}$,
making the independence on curve parameterization explicit. Its average is
\be
\langle W\rangle =\frac{Z[J_\mu]}{Z[0]}
\ee

\subsubsection{Overcoming basic problems}

We want to calculate $\langle W\rangle$. We face some problems:
\begin{itemize}
\item There is no weight for 
${\tilde A}_\mu (0)$. Luckily, ${\tilde J}_\mu (0)=0$ so
${\tilde J}_\mu$ is not coupled to this degree 
of freedom and we can forget about it, 
taking the integration over it to cancel between $Z[J_\mu]$ and $Z[0]$. 
Wilson loops appear to be ``infrared safe''.
\item The integral is also over $a_L(k)$ and is unbounded in that direction; as
mentioned, we can fix this by putting in a weight; this is done by multiplying the
integrand by
\be
e^{-\frac{1}{2a_0 g^2} \int \spc(\partial_\mu A_\mu )^2} =
e^{-\frac{1}{2a_0 g^2} \int \mom k^2 a_L(k)a_L(-k) }
\ee
The $a_0$ dependent terms cancel between $Z[J_\mu]$ and $Z[0]$ leaving $\langle W\rangle$
$a_0$ independent.
\item There is an infinite number of $k$-values, 
with $\sim |k|^{d-1}$ amplification as $k^2$
becomes large. The factorized path integral consists of an infinite number of finite
factors, and the product over all of them does not converge.  This is solved by introducing an ultraviolet cutoff $\Lambda$ and restricting the integration variables
by
\be
k^2 < \Lambda^2
\ee
Consequentially, the averaging process over 
the ${\tilde A}_\mu (k)$'s is now only sensitive 
to ${\tilde J}_\mu(k)$ with $k^2 < \Lambda^2$. The path integral does not affect
the dependence of $W$ on ${\tilde J}_\mu(k)$ with $k^2 > \Lambda^2$. 
\item $J_\mu(x)$ is a distribution: it is zero for $x\ne z(\tau)~ \forall \tau$ 
and infinite if for some $\tau$ we have $x=z(\tau)$. A Gaussian integral is 
bound to give (taking into account translational invariance) a Gaussian answer:
\be
e^{-\frac{1}{2}\int d^dx d^dy J_\mu(x) J_\nu(y) G_{\mu\nu}(x-y)}
\ee
Unless $G_{\mu\nu}(x-y)$ makes this irrelevant, we have to confront products
$J_\mu(x) J_\nu(y)$ at coinciding points $x$ and $y$. There is no general way to give
meaning to the product of distributions at the same point. This problem is now
naturally solved by simply changing our definition of the current, first in Fourier
space, by setting
\be
{\tilde J}_\mu (k)=0~~~{\rm for}~~ k^2 >\Lambda^2
\label{credef}
\ee
The new current is still conserved:
\be
J^\Lambda_\mu (x)=\int_{k^2 <\Lambda^2} \mom e^{-ikx} \int_0^l d\tau \frac{dz_\mu}{d\tau}
e^{ikz(\tau)}
\ee
$J^\Lambda_\mu(x)$ is no longer sharply localized at the curve; rather it is just peaked
at it and spread out over a distance $\delta \sim\frac{1}{\Lambda}$ in its vicinity.
Actually, because of 
my simple choice of a sharp momentum cutoff this localization
is not good enough for all cases, as the decay away from the curve is slow and there
is an oscillatory behavior of high frequency. 

In order to distinguish the curve's shape from that of an amorphous blob, we need
$\Lambda l\gg 1$. In principle, we want to rid ourselves of the dependence on the large
number $\Lambda l$ by taking it to infinity. Whatever we can make sense of in that
limit is a universal feature of the theory, since many different schemes of 
overcoming the problems listed would produce identical results when the 
cutoff is removed. We now turn to see what we can do; this depends on the dimension $d$.

\end{itemize}

\subsection{Circular loop}
{\it Exercises:}

Consider a loop given by a circle of radius $R=l/(2\pi)$ in the 1-2 plane.
Take $d=4$.
\begin{enumerate}
\item Calculate $\langle W \rangle$ for $J_\mu^\Lambda$. Answer has the form
$\langle W \rangle =e ^{-g^2 I/2}$ with
\be
I\sim - (\Lambda R)^2 \int_0^1 d\xi \log \xi {\cal J}_1^2 (R\Lambda \sqrt{\xi})
\ee
where ${\cal J}_n$ are Bessel functions of integral order $n$. 
\item What is the leading behavior of $I$ as $\Lambda R$ tends to infinity ?
Answer: 
\be
I=c_0 (\Lambda R) +{\rm lower~orders}
\ee
For obvious reasons the first term is known as the ``perimeter divergence''.
From the calculation it becomes obvious that for a curve of different shape, but same perimeter, this divergent term would be the same as for the circle. 
\item Consider now a change in the definition of the current $J_\mu^\Lambda$; instead
of~(\ref{credef}) we use a $J_\mu^{\Lambda^\prime}$ with $\Lambda^\prime < \Lambda$.
We still have $\Lambda^\prime R > 1$, say $\Lambda^\prime R =100$, but
$\Lambda\gg\Lambda^\prime$ and we now take $\Lambda R$ to infinity at 
$\Lambda^\prime R$ fixed at some number much larger than one. What happens
now ? Answer: The perimeter divergence disappears, and 
gets replaced by a finite perimeter term given by 
$\Lambda^\prime R$ times some number as the leading term in an expansion in
$\frac{1}{\Lambda^\prime R}$.  
\end{enumerate}

Take now $d=3$ and consider the same questions. The main difference is that now
the perimeter divergence is just logarithmic. Finally, take $d=2$. Now there are
no divergences.

The final conclusion is that the best place to 
start learning about Wilson loops is in two dimensions. 
There one can consider true, infinitely thin, curves. 
It may be possible to study loops also in 
higher dimensions, but this may require to
``fatten'' the curves by a finite amount, of relative order $\Lambda^\prime l$. 

\subsection{Geometric meaning of $W$}

\subsubsection{Matter}

In full QED, photons 
interact with Dirac fermions of charge $g$.
The path integration is extended to include the Grassmann variables 
$\bar\psi^\alpha (x),\psi^\alpha (x)$ and the action $S[A_\mu]$ is changed
by an additive term
\be
S_\psi = \int \spc \left [ \bar\psi(x)\gamma^\mu D_\mu \psi(x)+m\bar\psi(x)\psi(x)\right ];~~~D_\mu=\partial^x_\mu-iA_\mu (x)
\ee
[The Dirac indices $\alpha,\beta$ are silent and summed over. They
will not appear again.] 
This coupling preserves gauge invariance with $J_\mu$ being replaced by 
\be
\bar\psi\gamma^\mu\psi=J^\psi_\mu
\ee
Now, $\partial_\mu J^\psi_\mu$ is not zero for an arbitrary $\bar\psi,\psi$,
but, it is zero when the fields $\bar\psi,\psi$ both satisfy 
the extremum condition along with $A_\mu$, and an expansion of the entire
path integral in $g$ can be defined.
If one instead integrated out $\bar\psi,\psi$ first, the result $S^\psi_{eff}[A_\mu]$ is a gauge invariant addition to $S[A_\mu]$. 
If one instead integrated out $A_\mu$ first, the $a_L$ component
of that integration effectively puts a constraint on the remaining
$\bar\psi,\psi$ that enforces $\partial_\mu J^\psi_\mu=0$. If one
has added a weight factor for $a_L$, for any positive $a_0$ no constraint
on $\bar\psi,\psi$ gets generated and only a subset among all possible
$A_\mu,\bar\psi,\psi$ observables are deemed physical. That is the subset
of gauge invariant observables; they would be independent of the parameter
$a_0$. 

\subsubsection{Holonomy}

For $g=0$, $D_\mu=\partial^x_\mu$ and we know that $\partial^x_\mu$ generates
translations:
\be
e^{a_\mu \partial^x_\mu}\psi(x)=\psi(x+a)
\ee
In particular, for a closed curve we have
\be
e^{\oint dz_\mu \partial^x_\mu}\psi(x)=\psi(x)
\ee

Replacing $\partial^x_\mu$ by $D_\mu$ in the above equation gives
\be
e^{\oint dz_\mu [\partial^x_\mu-iA_\mu(z)] }\psi(x)=W^\ast \psi(x)
\ee
So, when $\psi(x)$ is ``transported'' round a closed curve it 
accumulates an additional phase factor (holonomy), given by the
Wilson loop operator associated with the curve.

\subsubsection{Infinitesimal loops}

For an infinitesimal loop,  bounding a flat surface element $\delta\sigma_{\mu\nu}$, one has
\be
W\approx 1+i\delta\sigma_{\mu\nu} F_{\mu\nu}
\ee
The role of $S[A_\mu]$ is seen to provide a bound on the fluctuation
in $W$ when the loop is very small. This works well in $d=2$, but, as we 
have seen, has problems in $d>3$.

\subsubsection{Some dimensional analysis}

For any $d$, $W$ is dimensionless. This implies that $A_\mu$ has dimension 
1 (this is mass dimension, the inverse of length) and therefore $F_{\mu\nu}$
has dimension 2. $S[A_\mu]$ is in the exponent, 
so also needs to be dimensionless. Therefore, the dimension of the coupling
$g^2$ is $4-d$. 

After averaging, one gets for a small loop, schematically, 
\be
1-\langle W\rangle \sim (\delta\sigma F)^2
\ee
Here we ignored the perimeter term, although it is divergent for $d>2$. 

Let $\delta l$ be the linear scale of the loop. The gauge action gives
\be
\frac{1}{g^2} (\delta l )^d F^2 \sim 1
\ee
Hence, with $\delta\sigma \sim \delta l^2$, 
\be
1-\langle W\rangle \sim (\delta l^2 F)^2 \sim g^2 (\delta l)^{4-d}
\label{coupling}
\ee

For a small loop this would go to zero as the loop shrinks, except at $d=4$, where
the theory is scale invariant in the present approximation. Once the
approximation is improved upon, scale invariance is lost even in $d=4$ and $g^2$ 
starts depending on $\delta l$, logarithmically for small $\delta l$. If the gauge
theory is nonabelian, one has $g^2(\delta l) \sim -1/\log(\Lambda_{SU(N)} \delta l) $ for $\Lambda_{SU(N)}\delta l
\to 0$ and one recovers $\langle W\rangle\to 0$ for $\delta l \to 0$ even at $d=4$. 

However, for $d>2$, the above effect is masked by the perimeter divergence.

\subsection{Summary of section}
\begin{itemize}
\item{Wilson loops associate a phase factor with parallel transport round a closed
curve}
\item{The gauge action is defined so as to suppress fluctuations  
of phase factors associated with infinitesimal loops 
away from identity}
\item{For small loops the phase factor is close to identity, but this is
masked by a perimeter divergence for $d>2$}
\item{For $d=4$ to behave similarly to $d=2,3$ we better focus on non-abelian gauge
theories, whose effective coupling tends to zero at short distances}
\end{itemize}

\section{The nonabelian holonomy}

\subsection{Definition}

The nonabelian gauge group has a Lie algebra ${\cal G}$ of real dimension $n$ and
generators labeled by $i=1,..,n$. We restrict our attention to ${\cal G}=su(N)$
with $n=N^2-1$. There are now $n$ vector fields $A^i_\mu (x)$. $su(N)$ is
compact and has a discrete infinity of irreducible finite dimensional, unitary
representations, in which the generators are represented by linear unitary operators
acting on finite dimensional Hilbert spaces. Each representation is labeled by $R$, 
the dimension of the associated Hilbert space by $d_R$ and the operators representing
the generators by $T^{(R)i}$. Choosing a basis, makes the generators traceless hermitian
matrices with entries denoted by $T^{(R)i}_{a,b},~a,b=1,..,d_R$. We adopt the convention
\be
\Tr T^{(R)i}T^{(R)j}\propto \delta^{ij}
\ee
so that the structure constants $C{ijk}$ are totally antisymmetric:
\be
[T^{(R)i},T^{(R)j}]=iC^{ijk}T^{(R)k}
\ee

For any $R$, we define a covariant derivative acting on a matter field $\psi^{(R)}_b (x)$
both on $x$ and on $b$:
\be
[D_\mu \psi ]_a(x)=[\partial_\mu^x \delta_{ab}-i 
T^{(R)j}_{ab} A^j_\mu (x) ]\psi_b (x)
\ee

Parallel transport round a closed curve has a holonomy given by the action of a unitary
matrix, representing a group element $W$ in the representation $R$:
\be
W_R (x) = P e^{i\oint dz_\mu A^j_\mu (z) T^{(R)j} }
\ee

\subsection{Gauge invariant content of holonomy}

The symbol $P$ indicates that the product 
is ordered round the closed path, starting
at $x$ and ending at the same point $x$. 
Ordering is necessary as the matrices
in the exponent, at different points on the path, 
do not commute with each other. 
A finite gauge transformation acts by an $x$-dependent 
group element $g(x)$ 
\be
W_R (x) \to g^{(R)} (x) W_R(x) g^{(R)\dagger} (x)
\ee
where $g^{(R)} (x)$ is the unitary matrix representing $g(x)$ in the irreducible representation $R$. Hence, the gauge invariant content of the holonomy is contained
in the infinite collection of numbers
\be
\chi_R (W) = \Tr W_R(x)
\ee
The notation indicates that the dependence on $x$ disappears, and these are indeed
numbers associated with the closed curve, without selecting any particular point
on it. $W$ is the abstract 
group element associated with parallel transport round the curve from
$x$ to $x$. 

Viewed as functions on group space, the $\chi_R (W)$ are a basis of the space of
all class functions on the group. Any class function is a smooth function of the
eigenvalues of 
\be
W_f (x) =  P e^{i\oint dz_\mu A^j_\mu (z) T^{(f)j} }
\ee
where $f$ denotes the fundamental representation; for $su(N)$, it is by hermitian, 
traceless $N\times N$ matrices. The eigenvalues are $N$ points on the unit circle
$e^{i\gamma_a}$, $\gamma_a$ real, constrained by $\prod_{a=0}^{N-1} e^{i\gamma_a}  =1$. $W_f$ is 
an $N\times N$ unitary matrix
of unit determinant. 

\subsection{A probabilistic view}

As is well known, one can write a gauge invariant Lagrangian for the fields
$A_\mu^i(x)$, but it is nonlinear. For a small loop one has
\be
W_R \sim 1+i\delta\sigma_{\mu\nu} F_{\mu\nu}^k T^{(R)k}
\ee
where
\be
F_{\mu\nu}^k =\partial_\mu A_\nu^k - \partial_\nu A_\mu^k +C^{ijk} A_\mu^i A_\nu^j
\ee
The action $S[A_\mu^i]$ still consists just of a suppression factor in the traces
of holonomies
round small loops:
\be
S[A_\mu^i] = \frac{1}{4g^2} \int\spc F_{\mu\nu}^k F_{\mu\nu}^k
\ee
The integrand is a class function, so the suppression is on the eigenvalues
of the small loop $W_f$. The action tries to make $W_f$ close to the $N\times N$
unit matrix for small loops. In other words, the action tries to make the holonomy
of tiny loops close to the identity in the group. 

One could imagine integrating out in the path integral all  $A_\mu^i(x)$ with
the constraint that, for a chosen fixed loop, its holonomy $W$ is kept fixed.
This would produce a positive function 
(we are setting to zero a potential second term in 
the action, that is possible at $d=4$, and which is imaginary) since the
integrand is positive. Normalizing, we would get a probability distribution
on the group, $P(W)$. Gauge invariance tells us that $P(W)$ 
must be a class function and that the measure of integration on $W$ must be the
Haar measure. $P(W)$ contains all the information needed to compute all
averages of the form
\be
\langle \chi_{R_1} (W) \chi_{R_2} (W) .... \rangle
\ee

Since $P(W)$ can be linearly expanded in $\chi_R(W)$ the coefficients in that expansion
hold all the information determining the above moments.

One can think about $S[A_\mu^i]$ as defining a highly peaked, 
``bare'' distribution $P_0 (W)$ for every tiny loop; the peaking is controlled by $g^2$.
The weight generated by the $P_0(W)$ for all tiny loops $W$ collectively 
produces the above $P(W)$ for a macroscopic loop. 

This picture has ignored the problems we have encountered in the 
case where the group is $U(1)$. These problems have not gone away, since for $g^2\to 0$
we have just $N^2-1$ non-interacting photons. However, there are no problems at $d=2$.

\subsection{$S[A_\mu^j]$ from $P_0(W)$}

Let us start with $P_0(W)$, from which we wish to 
derive $S[A_\mu^i]$ and $P(W)$.
To make an action out of $P_0(W)$ one needs to make sure that the collection
of infinitesimal loops that is selected includes every $A_\mu^i(x)$ (all
$x,\mu,i$). There is much freedom in the choice of $P_0(W)$ itself, as only
the behavior in the vicinity of $W=1$ matters. A very natural choice is provided
by the heat-kernel function: One writes down a diffusion equation for $W$ which
is consistent with the homogeneity of the group manifold:
\be
\frac{\partial}{\partial t} P_0 (W;t) \propto \nabla_W^2 P_0 (W;t)
\ee
$t\ge 0$ is dimensionless 
and $\nabla_W^2$ is the Laplacian on the group manifold: It is defined
by being the ordinary Laplacian in the tangent space at $W=1$ extended by conjugation
to the entire group. $\nabla_W^2$ maps class functions into class functions. 
The diffusion constant, which could be absorbed in $t$, 
is a number chosen by some convention, to make the formula
for $P_0(W,t)$, below, correct. 
The initial condition for the diffusion equation is taken as
\be
P_0 (W;0)=\delta_{\rm Haar} (W,1)
\ee
Thus, $P_0(W;t)$ will be a class function for all $t>0$. It is explicitly
given by
\be
P_0 (W;t)=\sum_R d_R \chi_R(W) e^{-\frac{t}{2N} C_2 (R) }
\ee
Here the sum over $R$ is over all irreducible representations of $SU(N)$ 
and the number $C_2(R)$ is the value of the 
quadratic Casimir operator on the representation $R$ in a specific 
convention for the normalization of the generators. 
For all $t\ge 0$ and $W$ we can view $P_0 (W;t)$ as a probability density
since $P_0 (W;t) >0$ and the diffusion equation preserves the initial normalization.  

The group elements $W\sim 1$ are identified by $W_f=e^{iH}\sim 1+ i H$, where
$H$ is $N\times N$, hermitian and traceless. For $t\ll 1$, the matrix norm of $H$
is small with high probability. We now see that we can identify
\be
H\sim \delta\sigma_{\mu\nu} F^j_{\mu\nu} T^{(f)j}
\label{hh}
\ee
where the small loop in question is the boundary of the little surface
element $\delta\sigma_{\mu\nu}$. We ensure that all $A^j_\mu(x)$ participate
by including all rotated and translated copies of $\delta\sigma_{\mu\nu}$,
which amounts to multiplying $P_0(W,t)$ factors over all $\mu>\nu$ and $x$. 

Taking $t<<1$ we can use the tangent space approximation to the diffusion equation
and we get the standard YM action. 

\subsection{Summary of section}

The holonomy is a geometric construct and is 
the central object of non-abelian gauge theory. 
Both the action and all physical observables are given by holonomies.

One can view the basic problem of Euclidean 
non-abelian gauge theory in $R^d$ as the calculation of the probability 
distributions of holonomies for arbitrary loops given the probability 
distribution of a complete set of holonomies of infinitesimal loops.
By a complete set, one means a set of loops whose holonomies determine
the connection $A_\mu^i(x)$, up to gauge transformations. In turn, the
$A_\mu^i(x)$ determine all holonomies and are the variables of integration in
the path integral.

\section{Two dimensions}

In two dimensions the calculation of $P(W)$ for a macroscopic, non-selfintersecting
loop, given $P_0(W)$, is straightforward. One needs to tile the area inside the loop
by an exactly fitting cover made out of elementary microscopic loops. 
So longs as such a cover exists, the shape of the
loop does not matter; just the total enclosed area affects $P(W)$.
The result is
\be
P(W)=P_0(W; \tau)
\label{prob}
\ee
where $\tau$ is the area ${\cal A}$ 
enclosed by the loop in units of $\lambda\equiv g^2 N$:
\be
\tau= \lambda {\cal A} \left ( 1+\frac{1}{N}\right )
\ee
Therefore, as we scale the loop up, its holonomy $W$ spreads from the vicinity
of unity for a small loop, to visiting the entire group evenly at $\tau=\infty$. 

\subsection{The Durhuus-Olesen~\cite{DO} non-analyticity at $N=\infty$}

I shall refer to this non-analyticity as a ``phase transition''. 
This phase transition occurs at $N\to\infty$ at fixed $\lambda{\cal A}$ when
$\tau$ increases through the value $\tau_c\equiv 4$. 

There are many ways to see the phase transition. I choose one of the simplest,
employing the average characteristic polynomial of $W_f$.

\subsubsection{Average characteristic polynomial~\cite{rburgers}}

{\it  Exercise}: Equation~(\ref{prob}) gives for a non-selfintersecting loop enclosing an area equal to $\tau$ in dimensionless units:
\be
\langle \chi_R(W(\tau)) \rangle= d_R e^{-\frac{\tau}{2N} C_2 (R)}
\ee

{\it Exercise}: Define 
\be
\psi^{(N)}(z,\tau) =\langle \det(z-W_f(\tau))\rangle
\ee
Show that the above polynomial in $z$ generates all the $\langle \chi_R(W(\tau)) \rangle$
with totally antisymmetric $R$ (single column Young pattern). 

{\it Exercise}: Define
\be
\phi^{(N)}(z,\tau)=\frac{1}{2}-
\frac{1}{N}\frac{z}{\psi^{(N)} (z,\tau)} 
 \frac{\partial \psi^{(N)} (z,\tau)}{\partial z} 
\ee
Prove that 
\be
\varphi^{(N)} (y,\tau)=\phi^{(N)}(-e^{y},\tau)
\ee
with real $y$, satisfies:
\be
\frac{\partial\varphi^{(N)}(y,\tau)}{\partial\tau}
+\varphi^{(N)} (y,\tau)\frac{\partial\varphi^{(N)}(y,\tau)}{\partial y}=
\frac{1}{2N} \frac{\partial^2\varphi^{(N)}(y,\tau)}{\partial y^2}
\label{burgers}
\ee
with initial condition
\be
\varphi^{(N)} (y,0)=-\frac{1}{2}\tanh\frac{y}{2}
\ee

{\it Exercise}: Show that the Burgers' equation in (\ref{burgers}), with the above
initial condition, 
produces, in the limit $N=\infty$, a shock at $y=0$ when $\tau$ reaches
the value $\tau=4$. For $\tau <4$ the solution is smooth in $y$. 

We see that $\frac{1}{2N}$ plays the role of viscosity in the Burgers equation.
The zero viscosity limit is singular, producing a ``breaking wave'' at $\tau=4$.
The singularity is absent at any finite $N$; hence the term ``infinite $N$ phase
transition''.

\begin{figure}
\centering
\includegraphics[width=10cm]{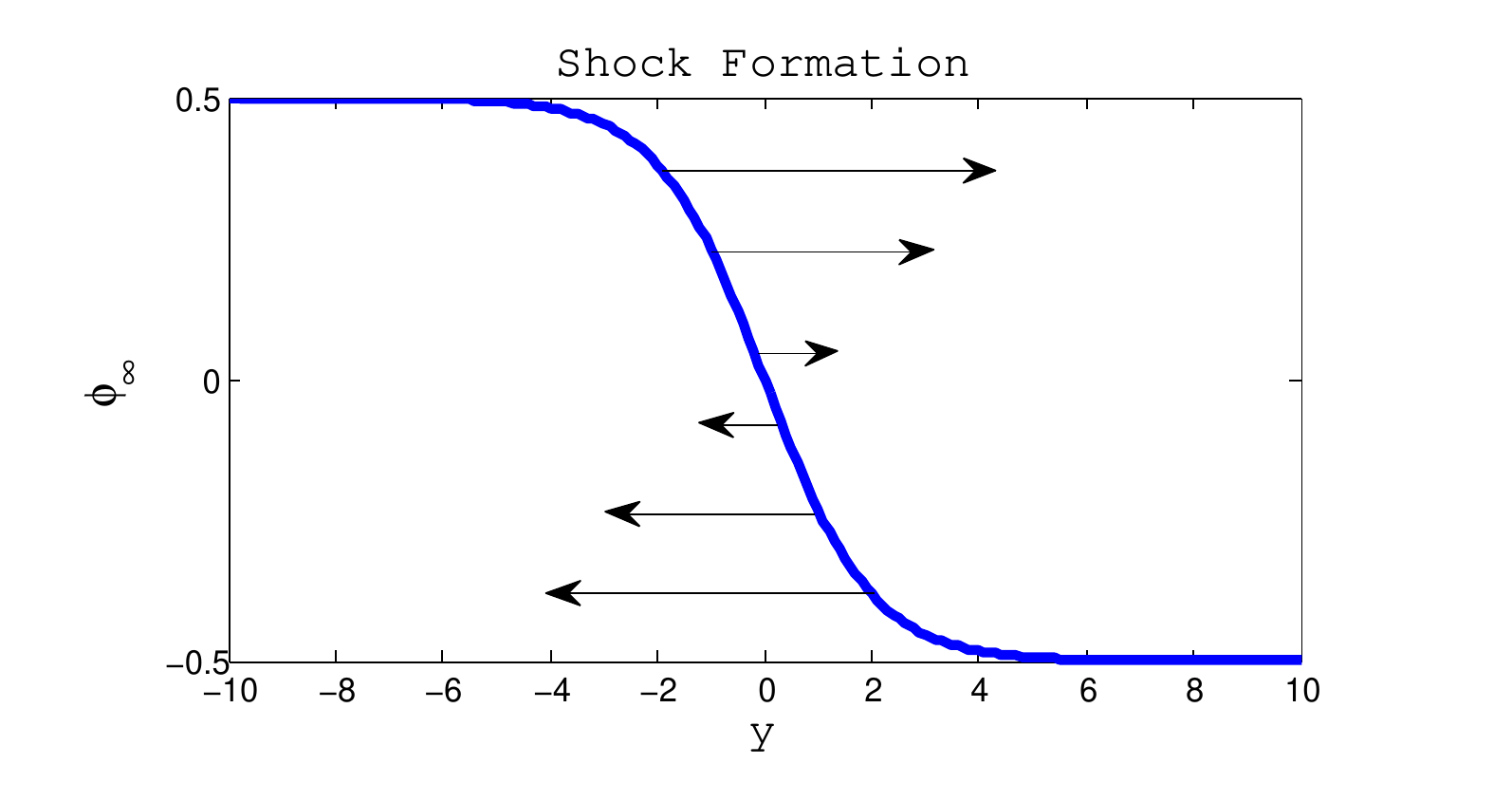}
\caption{The making of a shock.}
\label{shock}
\end{figure}

Note
that the critical value $\tau=4$ is determined by the initial condition and
not by the equation. Also note that deforming the initial condition would
not eliminate the appearance of the shock, except when the initial condition
is changed in a drastic manner. So, while the appearance of a shock is
a likely event, the exact value of $\tau$ where it happens varies from case 
to case.  

{\it Exercise}: Explain Figure~\ref{shock}, specifically, what the arrows mean
and why the caption is justified.

\subsubsection{Motion of zeros~\cite{twodnew}}

As a monic polynomial, $(-1)^N \psi^{(N)}(z,\tau)$ can be parametrized by its zeros.

{\it Exercise}: Prove that all the zeros $z_a(\tau),~a=0,..,N-1$ are on the unit circle. 

{\it Exercise}: Write $z_a (\tau) = e^{i\theta_a(\tau)}$ and work out the
equations of motion for the angles $\theta_a(\tau)$:
\be
\frac{d\theta_a}{ d\tau} =\frac{1}{2N} \sum_{b,\;b\ne a} \cot \frac{\theta_a -\theta_b}{2}
\ee
The initial condition is at a singular point:
\be
\theta_a(0)=0
\ee
Thus, the partial differential equation~(\ref{burgers}) 
with our specific initial condition can be reduced to a system of $N$ first order differential equations. 

Using these equations one can evaluate the motion of the zeros for $\tau \ll 1$ and for $\tau \gg 1$.

{\it Exercise}: Make the ansatz 
\be
\theta_a (\tau ) =2\eta_a\sqrt{\frac{\tau}{N}}
\ee
and solve the
equations of motion for the $\theta_a(\tau)$ to leading order in $\tau\ll 1$. 
The result is that the $\eta_a$ are the zeros of the Hermite polynomial of order $N$:
\be
H_N (\eta_a)=0,~~a=0,1,...,N-1
\ee

{\it Exercise}: Prove that for all $\tau > 0$
\be
\sum_{a=0}^{N-1} \theta_a (\tau) =0
\ee
and that $\left [\frac{N}{2}\right ]$ pairs of angles have the same absolute values and opposite signs. Also, show that if $N$ is odd there is one angle that stays at $0$ for
all $\tau$. 

{\it Exercise}: Find the asymptotic behavior of the zeros as $\tau\to\infty$.
The result is 
\be
\theta_a (\tau=\infty)  = \frac{2\pi}{N}\left ( a-\frac{N-1}{2}\right ) \equiv \Theta_a
\ee
At infinite $\tau$ the zeros are uniformly spread round the unit circle. 

{\it Exercise}: How do the zeros approach their $\tau=\infty$ values ?
Set $\theta_a (\tau)=\Theta_a +\delta \theta_a (\tau)$. The result, to leading order, is
\be
\delta\theta_a(\tau) \sim - 2e^{-\frac{\tau}{2N} (N-1)} \sin\Theta_a
\ee
Note that the zeros with $a\sim\frac{N}{4},\frac{3 N}{4}$ 
move the fastest at large $\tau$.

{\it Exercise}: Let $N\gg 1$ and fixed. Calculate the asymptotic behavior of the
pair of zeros closest to $-1$ on the unit circle at $\tau=4$. The answer is
\be
z_M \sim -\exp \left [ \pm \frac{3.7i}{ N^{\frac{3}{4}}} \right ]
\ee
The number 3.7 is an approximation. 

{\it Exercise}: Let $N\gg 1$ and fixed. 
Let $\frac{\tau}{4}=1+\frac{\alpha}{N^\nu}$.
Find $\nu$ such that $z_M (\tau)$ has a finite 
nontrivial dependence on $\alpha$
of the form 
\be
z_M \sim -\exp \left [ \pm \frac{f(\alpha) i}{ N^{\frac{3}{4}}} \right ]
\ee
as $N\to\infty$, with $f(0)\approx 3.7$. The answer is $\nu=1/2$. 

We see that the motion of the extremal zeros becomes nontrivial in the limit
$N\to \infty$ if we express them in terms of specific scaling variables. The associated exponents of $N$ are $1/2$ and
$3/4$. 

\subsection{Zeros and eigenvalues}

At large enough $N$, the set of zeros $z_a(\tau)$ is a good approximation to the
set of most likely values of the eigenvalues of the fluctuating matrix $W_f$:

Too see this we need to compute the single eigenvalue density $\rho_N (\theta;W)$.
\be
\rho_N(\theta;W) = \frac{1}{N} \sum_a \langle \delta_{2\pi} (\theta - \gamma_a(W))
\equiv\rho_N (\theta , \tau) \rangle
\ee
where the the eigenvalues of an instant of $W_f$ are $e^{i\gamma_a}, a=0,..,N-1$. 

{\it Exercise}: Compute $\rho_N(\theta ,\tau)$. Start by expanding 
$\det(1+uW_f)/\det(1-vW_f)$ in characters, then take the average and next 
study the limit $u\to -v$.

\begin{figure}
  \begin{tabular}{l@{\hspace*{20mm}}r}
    \includegraphics[width=0.4\textwidth]{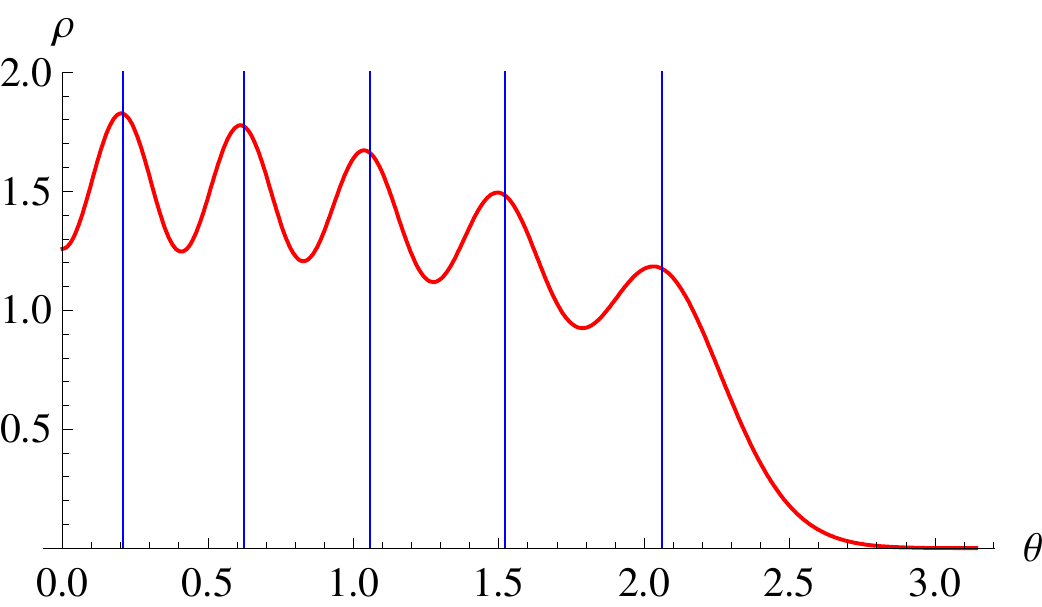} &
    \includegraphics[width=0.4\textwidth]{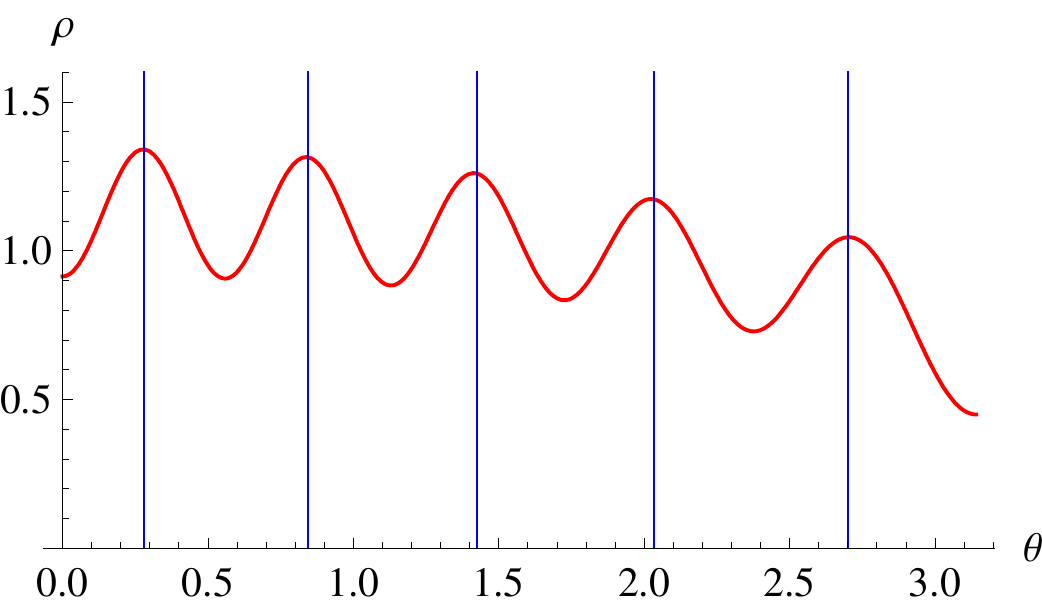} \\
    \includegraphics[width=0.4\textwidth]{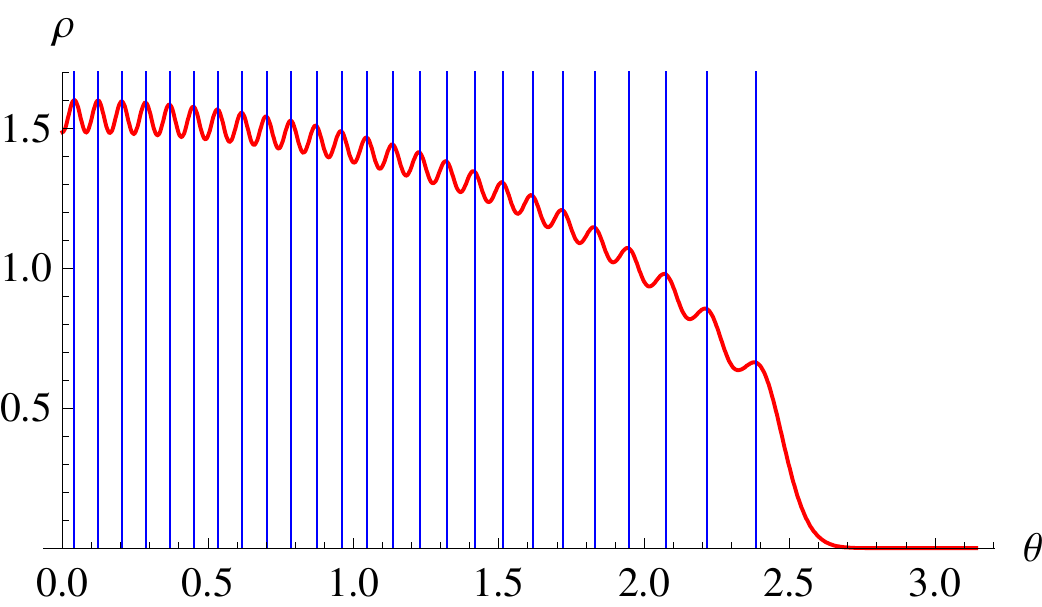} &
    \includegraphics[width=0.4\textwidth]{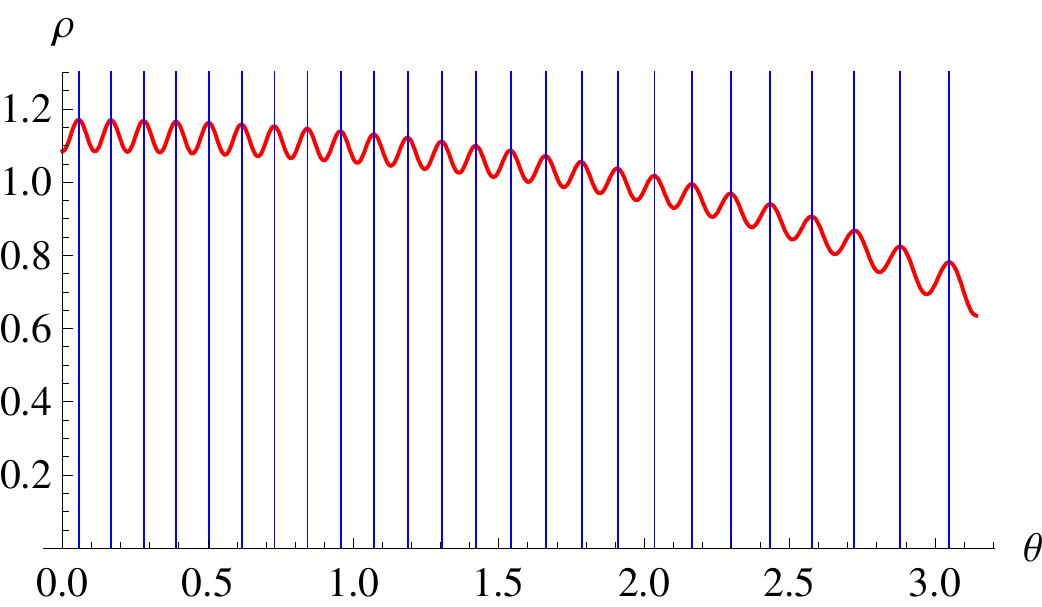} 
  \end{tabular}
  \caption{Plots of the density $\rho_N (\theta )$ (oscillatory
    curve) together with the positions of the angles of the
    zeros $\theta_a$ (vertical lines) for $\tau < 4$ (left)
    and $\tau > 4$ (right), $N=10$ (top), and $N=50$~(bottom).}
  \label{figRhoTrueZeros}
\end{figure}

At $N=\infty$, as first shown by Durhuus and Olesen~\cite{DO}, $\rho_\infty(\theta,\tau)$ has
a gap centered at $\theta=\pm\pi$ for $\tau < 4$, which closes for $\tau >4 $.
The exact formula for finite $N$ shows that there are $N$ oscillations modulating
the $N=\infty$ form and, when the $N=\infty$ case has a gap, the finite $N$ density
has a non-vanishing tail, exponentially suppressed with $N$ in the gap region.

The peaks of the oscillations can be interpreted as the 
most likely locations for the
fluctuating eigenvalues of $M_f$, with phases 
$\gamma_a$. The plots in Figure \ref{figRhoTrueZeros} show how they compare to
the locations of the zeros $z_a$. We may loosely refer to the zeros $z_a$'s
as the eigenvalues of $W_f$ or even $W$.

The behavior of $\rho_N(\theta,\tau)$ close to critical $\tau$ 
and for $\theta$ close to $\pm\pi$  is universal at large $N$; this statement
can be made precise by defining ``close'' for the two parameters 
$\theta$ and $\tau$ with the help of
two scaling variables $\xi,\alpha$ and the associated critical exponents of $N$,
$3/4,1/2$ respectively. 
The extremal zeros $z_M$ reside in the critical 
regime of $\theta$ when $\tau$ is also in the critical regime.

\subsection{Summary of section}

For a small loop, the eigenvalues of $W_f$ are all concentrated around unity.
As the loop expands the eigenvalues expand round the unit circle until they
cover it completely. Very large Wilson loops have eigenvalues randomly
distribute round the unit circle. There is a crossover between the two regimes.
This crossover can be seen in various ways: The evolution of the average
characteristic polynomial as a function of $y$, where $z=-e^y$ with $y$-real, 
shows a sharp behavior at $y=0$ when the size of the loop is in the crossover
regime. The extremal eigenvalues close the gap in the crossover regime. 
When taking $N\to\infty$ the crossover turns into a phase transition, and
for large enough $N$ there is a regime of universal behavior. 

The main qualitative change between the perturbative regime for small loops and
the non-perturbative regime for large loops is that the eigenvalues cover just 
a small arc round $z=1$ in the former and the entire unit circle in the latter.
The crossover is when the entire unit circle first gets covered. The zeros of
the average characteristic polynomial go as $e^{i\sqrt{\tau} \eta_a}$ for small
$\tau$ and like $\exp\left\{i\left[\Theta_a -2 e^{-\sigma \tau}\sin\Theta_a\right ]\right\}$ for
large $\tau$, where $\sigma$ is the fundamental string tension 
and the $\Theta_a$ cover the circle uniformly. 

Our main hypothesis is that this behavior also holds in $d=3,4$ once holonomies
are properly defined, eliminating the perimeter divergences. If this is true,
one can hope that in $d=4$ the regimes of confinement and perturbation theory
are separated at large $N$ by a narrow regime in which one has universal behavior.
Further, I hope that this will allow to actually match perturbative to non-perturbative behavior in $d=4$ leading to a way to estimate the string tension
in four dimensions in terms of a perturbatively defined scale, $\Lambda_{SU(N)}$. 

In short,
the holonomy evolves with scale from exploring an 
arc round $z=1$ to filling the entire unit circle and the crossover between the
two realms narrows as $N\to\infty$. For $N\gg 1$ the crossover can be
described universality, using a simple random matrix theory effective 
description of $W_f$. 

\section{Higher dimensions}

In $d>2$ our hypothesis needs to be tested numerically. 
I shall set up first the general (that is, for $d=3$ and $d=4$) strategy for carrying out a test.

This strategy has been implemented fully in $d=3$ and partially in $d=4$. 
$d=3$ is cheaper in computer time than $d=4$ and there is one 
additional simplification: there are no ``corner divergences''.
We have seen that at $d>2$ there is a perimeter divergence. Only the 
existence of a piecewise continuous 
tangent to the curve enters in determining
the term of highest degree of divergence. 
If the curve has kinks, that is, it's tangent
has isolated  discontinuities, sub-leading logarithmic 
divergences localized at the kinks
develop. At $d=3$ the perimeter divergence is logarithmic, so kinks produce 
no sub-leading divergences. For numerical tests one employs 
typically hyper-cubic lattices and closed contours on the lattice must have
kinks. In $d=2$ the curve only needs to be continuous, even the existence
of a tangent is not required.

\subsection{Smearing~\cite{ourjhep}}

We need to eliminate the perimeter divergence. As anticipated, we do this
by ``fattening'' the loop; this has a major effect on small loops and a limited
effect on large ones. We need to extend the ideas from the abelian case 
to the non-abelian one. We introduce an extra parameter $\rho \ge 0$ 
[do not confuse the parameter $\rho$ with the eigenvalue density 
$\rho_N(\theta;W)$], and make
the $A_\mu^j (x)$ variables, that enter the definition of the Wilson loop,
$\rho$ dependent. In this sense, one may think about $\rho$ as adding a 
fifth dimension, taking $R^d$ to $R^d\times R_+$. But, it is only the
Wilson loop that feels $\rho$. The pure Yang-Mills theory being probed
lives still at $\rho=0$. We define new variables $A_\mu^j (x,\rho)$ by
a diffusion-like equation in $\rho$:
\be
\frac{\partial A_\nu^j (x,\rho)}{\partial \rho} = 
\left [ D_\mu F_{\mu\nu}(x,\rho)\right ]^j
\ee
where $D_\mu$ is the covariant derivative in the adjoint representation 
with respect to $A_\mu^j (x,\rho)$ and
\be
F_{\mu\nu}^j (x,\rho)=\partial_\mu A_\nu^j (x,\rho) - \partial_\nu A_\mu^j (x,\rho)
+ C^{ikj}A_\mu^i (x,\rho) A_\nu^k (x,\rho)
\ee

The equation is supplemented by the initial condition
\be
A_\mu^j (x,0)= A_\mu^j (x)
\ee
where the 
$A_\mu^j (x)$ are the fields in $S[A_\mu^j]$ and are the 
integration variables of the path integral. 

The smeared Wilson loop is defined by
\be
W_R=P\exp[\oint dz_\mu A_\mu^j (z,\rho) T^{(R)j} ]
\ee
Although we do not do this here, it is appealing to think of replacing
the constant $\rho$ above by a function $\rho(\tau)$, where the loop is
described by $z_\mu (\tau)$. 
It is important to realize that the behavior under gauge transformations 
in $R^d$ is preserved. 

How does this fatten the loop ? At leading order in perturbation theory
we just have $N^2-1$ photons and the equations linearize. In Fourier 
space we have 
\be
\frac{\partial {\tilde A}_\mu^j (k,\rho)}{\partial \rho} = 
-(k^2\delta_{\mu\nu}-k_\mu k_\nu) A_\nu^j (k,\rho)
\label{smear}
\ee
with
\be
 A_\mu^j (k,0)= A_\mu^j (k)
\ee
The longitudinal component of $A_\mu^j(k,\rho)$ does not vary with $\rho$ but
it does no appear in the action and does not couple to the Wilson loop either.
Each of the 3 transverse components, for each $j=1,N^2-1$, goes as
\be
a^j_\perp (k,\rho) = e^{-\rho k^2} a^j_\perp (k)
\ee
Hence, the coupling of the momentum modes of $A_\mu^j(x)$ with
$k^2\rho \gg 1$ to the closed curve 
is exponentially weakened. We have not literally fattened the
curve -- it stays a mathematical one dimensional curve, 
but the path integral integration variables do not
resolve the curve well over distance smaller than $\sqrt{\rho}$. 

To avoid losing the dependence on shape for small curves altogether, we can
make $\rho$ depend on the scale of the loop, perhaps most fittingly on its
perimeter (assuming, say, that the loops is an arbitrary, 
but smooth and small, deformation of a circle).
For large curves there is no danger of losing sight of the loop shape so we
imagine choosing
\be
\rho = \frac{l^2}{[l\Lambda^\prime]^2 + c}
\ee
where $c$ is some large number, say $c=20$. 

The above construction is quite ad-hoc, but we hope that it will allow
us to see the universal large $N$ crossover we expect in analogy 
with the $d=2$ case and that the latter will not depend on the details of 
our ``fattening'' procedure.

The main advantage of the ``fattening'' procedure we chose is that it can
be easily extended to lattice gauge theory, so we can test our hypothesis 
outside perturbation theory. 
The point is that the right hand side of (\ref{smear}) 
is the variation of the classical action, and therefore has a natural lattice 
counterpart, where the continuum integration
variables $A_\mu^j $ are replaced by the lattice integration
variables $U_\mu(x)$ [a $SU(N)$ $N\times N$ matrix attached to the 
link  connecting the grid vertex at $x$ to its nearest neighbor
in the $\mu$ direction] 
and the continuum action by the lattice action. The lattice
action itself is literally constructed 
as a product over all identical, sharply
peaked, probabilities for the holonomies going round all the 
most elementary loops on the hypercubic lattice [going round 
small two dimensional flat squares called plaquettes], 
so the basic  
concept on the lattice is the same as in the continuum. In practice, 
the differential equation in $\rho$
is discretized in the lattice context because it
has to be solved numerically, being nonlinear even in the continuum; 
there are various variants of doing this
and the precise 
procedures are known in Lattice Gauge Theory as ``smearing'' methods. 

Note that the smearing method will also eliminate ``corner divergences''
so it is adequate also for $d=4$ dimensions, not just $d=3$. 

\subsection{The lattice test}

The lattice is taken as a $d$-dimensional hypercubic grid of finite extend in
each direction and connected as a torus in each direction. On each lattice
link one has the link variable $U_\mu(x)$ described above. The links are all
given an orientation and parallel transport is now taking a matter field
from one site to the neighboring one by multiplication by  $U_\mu(x)$ if the
link is traversed in the positive direction, or $U^\dagger_\mu (x)$ if the
link is traversed in the negative direction. The collection of all $U_\mu(x),~\forall x,\mu$ can be 
thought of as defining the $d$ operators $e^{D_\mu}$. Transporting from site to site
one can parallel transport along any path made out of consecutive links. 
The parallel transporters round the elementary plaquettes, 
identified by a site $x$ and directions $\mu,\nu$ are denoted $U_p$ where
$p$ identifies the plaquette. The lattice action is
\be
S[U_\mu (x)]\propto\sum_p \Tr [1-U_p (x)]
\ee
One uses the notation $\beta$ for $\frac{1}{g^2}$ and $b$ for $\frac{1}{Ng^2}$.

Given a set $\{U_\mu(x)\}$, one produces a set of smeared $SU(N)$ matrices 
$\{ U_\mu(x,\rho)\}$ from which one constructs smeared Wilson loop matrices
$W_f$ using the smeared link matrices as elementary parallel transporters. 
For definiteness, we shall only consider square loops, in a lattice plane.
These loops have a side consisting of $L$ links, where $L$ is an integer.

There are no dimensional parameters on the lattice; dimensional analysis in the continuum is just an aid in carrying out various scaling transformations. There
is no intrinsic meaning to dimensional quantities in Field Theory. 

The total number of sites of the lattice is denoted by the integer $V$ and is 
referred to as the ``volume''. One needs to make $V$ so large that the results
obtained can be extrapolated to the limit $V\to\infty$. This is referred to
as the thermodynamic limit and, excepting computational cost, causes few problems
of principle in pure non-abelian gauge theories because they have a mass gap. 

On the lattice, everything that was said before, which did not quite hold
because of various ultraviolet problems, now holds in a precise sense. We do
have a well defined problem in classical 
statistical mechanics. The continuum results
are obtained by taking the continuum limit. One must make sure that 
one only poses questions
that have finite non-trivial answers in the continuum. 
The ideology behind this is the concept of
Field Theory universality, which says that physical continuum answers
will not depend on any details of the lattice definitions and will admit 
an asymptotic expansion that coincides with the asymptotic expansion in the 
continuum, defined by renormalized perturbation theory via Feynman diagrams.

{\it Exercise}: Show that the above works in $d=2$. Smearing is not needed.

\subsubsection{The ingredients of a lattice simulation}

The heart of the process is a code, 
a computer program which implements a Monte Carlo
process that generates sets $\{U_\mu (x)\}$, ``configurations'', with
probability
\be
e^{-S[U_\mu(x)]}
\ee
The basic idea behind these codes is that they implement a Markov chain made
out of simple steps to which one inputs a stream of pseudo-random numbers
generated in one of the better ways known. It is assumed that the numerical 
errors induced by the stream being pseudo-random rather than truly random are
far below the statistical errors induced by sampling. Sampling is used to
estimate averages of the interesting observables, Wilson loops in our case. 
One is restricted to finite samples and one wants to get good estimates
at minimal cost of computation time.  

Thus, one has a computational method of finding out the numerical value 
of the average $\langle \chi_R (W_f)\rangle$ for any lattice loop [in our case, 
square loops of side $L$] to some accuracy. 

There are several parameters one can set: $N$, the group size; 
$R$, the representation; $b$, the lattice
coupling; $L$, the loop size and $\rho$, the smearing parameter. There are
more parameters one is free to set in the code, 
that impact efficiency, controlling several aspects
of the algorithms and the precise form of the Markov chain. We shall assume
they have been fixed and are not changed during the simulation. 

The result of the simulation would then be a table of numbers for 
$\langle \chi_R(W)\rangle$ one entry for each set $(N,R,b,V,L,\rho)$.
If one intends to also test for Field Theoretical universality, one
can vary over different $P_0 (W)$ functions, changing the action $S[U_\mu(x)]$.

\subsubsection{Analysis}

Once the tables of numbers have been produced (the ``data'') one needs to
analyze this data. 
One part of the analysis is purely statistical and determines
the reliability of the data, expressed in one of the standard ways one quantifies 
statistical uncertainty. Loosely speaking, this stage determines the ``errors''. 

Here I shall only focus on the second stage of the analysis, the extraction
of estimates in the continuum limit. Since this stage may involve the data
in very nonlinear ways, another step 
of statistical analysis is needed, 
to determine how the errors propagate to the final 
answers. 

I shall not describe any of the error analysis, nor the algorithm used to 
generate the data.

\subsubsection{How our hypothesis is tested~\cite{threed}}

To get to the continuum limit one needs to take $b\to\infty$ and $V,L\to\infty$
at fixed $b$ in a correlated manner. One tries to make $V$ large enough from the
start, so that one can forget about it. This is easier as $N$ gets larger because
of a phenomenon known as ``large $N$ reduction''. 
Large $N$ reduction implies that increasing $N$ reduces 
the size of finite volume corrections. 

To determine how the limits
$L,b\to \infty$ have to be correlated  one keeps a selected physical
quantity fixed. 
For example, one could 
calculate the lattice string tension $\Sigma(b,N)$ 
\be
\Sigma(b,N)=-\lim_{L\to\infty} \frac{\log \langle W_f (L) \rangle }{L^2}
\label{tension}
\ee

One then defines a scale, a dimensionless number (dropping the 
explicit mention of the dependence on $N$) $\Delta(b)$, by
\be
\Delta(b)=\sqrt{\Sigma(b,N)}
\ee
It is very common to assign a dimensional number to $\Delta(b)$ 
called the ``lattice spacing'' $a(b)$ by
\be
a(b)=\frac{\Delta(b)}{440\;MeV } 
\ee
where I just picked some approximate value for the QCD string 
tension as $(440\;MeV)^2$. For large $L$
\be
\log \langle W_f (L) \rangle \sim -\Sigma(b,N) L^2 
\ee
and we write
\be
\Sigma(b,N) L^2 = [L a(b)]^2 \left [ \frac{\Delta(b)}{a(b)}\right ]^2
\ee
Both factors on the right hand side are 
physical dimensional quantities in the continuum limit.

It is a fact that as $b\to \infty$,
$\Sigma(b,N)$ goes to zero, so $a(b)$ goes to zero as $b\to\infty$.
This fact is relatively easy to understand. What is less trivial to establish
is that the left hand side of eq. (\ref{tension}) is nonzero. The latter
fact says that there is confinement on the lattice. To actually establish
confinement in continuum one needs to also show that $a(b)$ goes to zero
in the precise way predicted by continuum perturbation theory.

With $a(b)$ established, the continuum limit is defined by taking
$L\to \infty$ in such a manner that $l$, defined by
\be
l=L a(b)
\label{La}
\ee
is kept finite at a physical distance. That the distance is ``physical'' 
is usually expressed by converting $\Delta(b)$ to $a(b)$ and attaching units
to $l$. I shall use this language from now on, but, remember, the entire
calculation has no idea what an $MeV$ is. 

For each $b$ we compute, with $L=l/a(b)$, at $l$ held fixed, the eigenvalue
density
\be
\rho_N(\theta; W_f(L,\rho(L(b),b))
\ee
Here
\be
\rho(L,b)=\frac{l^2}{a^2(b) [l^2 + c a^2(b)]}=\frac{L^2}{a^2(b)[L^2+c]}
\ee
where $c$ is a pure number, say $20$, that is, significantly larger than 1.

In principle, we would like now to carry out the following ordered set of steps:
\begin{enumerate}
\item Select an $l$ and a $N$.
\item Select a sequence of increasing $b$ values, for which $V$ is
large enough to be assumed infinite, 
paired with a sequence of $L$'s such that eq.~(\ref{La}) is
obeyed.
\item For each pair $(b,L(b))$ in the above sequence calculate the
eigenvalue density $\rho_N \left (\theta; W_f [L(b),\rho(L(b),b)] \right )$
\item Extrapolate $b\to\infty$ and determine the continuum eigenvalue
density $\rho_N(\theta,l)$. 
\item Repeat the steps above, still keeping $l$ fixed, but increasing $N$
\item Obtain, by extrapolation to $N=\infty$, $\rho_\infty (\theta,l)$.
\item Repeat the above steps, now varying $l$. The range of $l$ one should use
should include the transition point $l_c$
\end{enumerate}

The hypothesis will be supported by the test if we find a finite $l_c$
equal to $c\; GeV^{-1}$, where is $c$ is a pure number 
of order 1, say between 0.1 and 10. The part of the hypothesis that is
being tested is just the existence of the transition, defined as the
demarcation point separating a $\rho_\infty (\theta ,l)$ with a gap around $\theta=\pi$ for 
$l < l_c$ from a $\rho_\infty (\theta, l)$ with no gap at $\theta=\pi$ for
$l>l_c$.

Next we need to devise a test of large $N$ universality. This is done
as follows:

We start by defining the lattice average characteristic polynomial:

\be
O_N(b,L,\rho(L,b) )=\langle\det\left ( e^{\frac{y}{2}} + e^{-\frac{y}{2}} W_f (L)\right ) \rangle
\ee

We would like to take $b\to\infty$. We already know from the
first part of the hypothesis test 
where the phase transition is expected to occur
on the lattice, and have checked that the critical size of the
smeared loop has a reasonable continuum limit.

We are
interested in the region $y\sim 0$, $N\to\infty$ and $l\sim l_c$.  
We now proceed by ordering the $N\to\infty$
and $b\to\infty$ limits differently.  
This amounts to making an extra 
assumption: We assume that the infinite $N$ limit of the
continuum limit is the same as the infinite $b$ limit of the infinite $N$
limit of the lattice theory. So, at fixed $b$ (which fixed $L$ and $\rho$),
we take $N$ to infinity first. The successful result of the first
part of the test is the main reason why we are willing to add this assumption.
The reversal of limits simplifies the procedure significantly. 

To test for the universality component of the 
hypothesis we need to identify at fixed $L$ and at infinite $N$ 
a critical coupling $b_c(L)$ where the spectrum of $W_f(L)$ just closes its
spectral gap at eigenvalues equal to -1. 
$a(b)$ is monotonically decreasing as $b$ increases. So, the physical size
of the loop $l=La(b)$ shrinks as $b$ increases. Dilating the loop corresponds
to decreasing $b$. Varying $b$ at fixed $L$ will take us through the transition.
Looking at plots of $\rho_N(\theta)$ for
$W_f$, one can observe this. A more quantitative method is as follows:

Take the numbers $O_N(b,L,\rho(L,b))$ and expand in $y$ around $y=0$.
I suppress the dependence on $L$, as it is held fixed.
\be
O_N(y,b)=C_0(b,N)+C_1(b,N) y^2 +C_2 (b,N) y^4+...
\ee
Define
\be
\Omega(b,N) = \frac{ C_0(b,N) C_2(b,N)}{C_1^2(b,N)}. 
\ee
If $N$ is large enough, 
and if we set $b=b_c(L,N=\infty)$ we should get a value close to the number
$\Omega (b_c,\infty)$. We define an approximation to $b_c(L,N=\infty)$, 
$b_c(L,N)$, by the equation:
\be
\Omega(b_c(L,N),N) = 
\frac{\Gamma(\frac{5}{4}) \Gamma(\frac{1}{4})}{6 \Gamma^2(\frac{3}{4})} = 
\frac{\Gamma^4(\frac{1}{4})}{48\pi^2}= 0.364739936
\label{method1}
\ee
The number above is taken from the same quantity defined for $d=2$.

{\it Exercise}: Calculate $\Omega (b=b_c,\infty)$ in $d=2$.

Now the objective is to establish that indeed the limit 
\be
b_c(L)=\lim_{N\to\infty} b_c(L,N)
\ee
exits. We are now in a position to find the critical physical size $l_c$
again and make sure we get the same result as in the opposite order of
limits. For this we repeat the above procedure for several values of $L$. 
Then, we first invert the function $b_c (L)$:
\be
b_c (L_c (b))=b
\ee
Actually, the set of values $b_c(L)$ is discrete, since $L$ only takes 
discrete values.
The function $L_c(b)$ defined above is a continuous interpolation of this
set of discrete values. We now want to see that the following limit
exists:
\be
l_c=\lim_{b\to\infty} L_c(b) a(b)
\ee
We find the same $l_c$ as before, with the same characteristics. 
To test for large $N$ universality 
we need to go to a regime of asymptotically large values of $N$. 
We do this again first, before taking the continuum limit, 
assuming that this reversal of limits is allowed.
 
We already know that $O_N(y,b)$ will 
exhibit critical behavior at $b=b_c(L)$ and $y=0$
as $N\rightarrow\infty$. There, it will obey 
large $N$ universality if we can show that there exists a non-universal normalization factor, ${\cal N}(b,N,L)$, 
smooth in $b$ at $b=b_c$, such that: 
\begin{eqnarray}
&\lim_{N\rightarrow\infty} {\cal N}(b,N,L)
O_N\left(y=
\left(\frac{4}{3N^3}\right)^{\frac{1}{4}}\frac{\xi}{a_1(L)},
b=b_c(L)\left [ 1 +\frac{\alpha}{\sqrt{3N}a_2(L)}\right ] \right)=\cr& 
\zeta(\xi,\alpha)
\end{eqnarray}
${\cal N}(b,N,L)$ is a normalization factor we have to include just in order to
get a finite answer.  $a_1(L)$ and $a_2(L)$ are $N$-independent 
numbers, non-universal parameters.

The universal function comes from carrying out the above calculation in $d=2$
and is given by:
\be
\zeta(\xi,\alpha)=\int_{-\infty}^{\infty} du e^{-u^4-\alpha u^2+\xi u }
\ee

{\it Exercise}: Verify the above and find out what $a_1,a_2$ are in $d=2$. 

{\it Exercise}: Calculate the function $\Omega(\alpha)$ in $d=2$. 

The essential ingredient that needs to be verified now are the critical exponents
$1/2,3/4$. 

The parameter $a_2(L,N)$ is
obtained by first setting
\be
b=b_c(L,N)\left[ 1 + \frac{\alpha}{\sqrt{3N}a_2(L,N)}\right],
\label{bscale}
\ee
where $b_c(L,N)$ has been defined above.  Next we write 
the derivative of $\Omega$ with
respect to $\alpha$ at $\alpha=0$, which is at the critical size, 
and set the result equal to the corresponding 
universal number in the large $N$ limit. 
\newpage 
\begin{eqnarray}
&\left.\frac{d\Omega(b,N)}{d\alpha} \right |_{\alpha=0} =
\left.\frac{1}{a_2(L,N)\sqrt{3N}} \frac{d\Omega}{db}\right |_{b=b_c(L,N)}
= \frac{\Gamma^2(\frac{1}{4})}{6\sqrt{2}\pi}
\left( \frac{\Gamma^4(\frac{1}{4})}{16\pi^2} -1\right)
=\cr & 0.0464609668
\label{a2ex}
\end{eqnarray}

{\it Exercise}: Verify the above formula.

$\frac{d\Omega}{db}$ would be close to maximal at $b=b_c$; hence 
$\frac{d\Omega}{db}$ varies relatively little as $b$ stays close to $b_c$. Since 
$b_c$ is not known to infinite accuracy the reduced sensitivity on the exact 
value of $b_c$ is an advantage which motivates this choice for defining 
$a_2(L,N)$.   
Unlike $b_c(L,N)$, the definition of $a_2(L,N)$ involves 
going into the large $N$ critical regime around $b_c(L,\infty)$ and
non-standard powers of $N$ come in.  
If our hypothesis is correct, $a_2(L,N)$ has to approach 
a finite non-vanishing 
limit
$a_2(L,\infty)\equiv a_2(L)$. This tests for the exponent $1/2$.

One also
expects that the limit be approached as a power series in $\frac{1}{\sqrt{N}}$. 
This is borne out by the data.

To test for the other exponent, we first set
\be
y=
\left(\frac{4}{3N^3}\right)^{\frac{1}{4}}\frac{\xi}{a_1(L,N)}
\ee
and then form a ratio whose value at infinite $N$ is again
a universal number we can easily compute. 
\be
\sqrt{\frac{4}{3N^3}} \frac{1}{a_1^2(L,N)}
\frac{C_1(b_c(L,N),N)}{C_0(b_c(L,N),N)}
= \frac{\pi}{\sqrt{2}\Gamma^2(\frac{1}{4})}=0.16899456
\label{a1ex}
\ee
This relation defines $a_1(L,N)$. 

{\it Exercise}: Verify the above formula.

Similarly to $a_2(L,N)$, the definition of $a_1(L,N)$ involves 
going into the large $N$ critical regime around $b_c(L,\infty)$. 
Consequently, we expect $a_1(L,N)$ to also have a finite limit
$a_1(L,\infty)\equiv a_1(L)$. This tests for the exponent $3/4$.
The approach to the limit is again by a power series in $\frac{1}{\sqrt{N}}$.

Finally, to make sure that all this survives the continuum limit
one needs to take $L\to\infty$ at fixed physical size $l=La(b_c(L))$, 
as explained earlier.

\subsection{Summary of numerical work}

A complete numerical test has been carried out in $d=3$. The results are
consistent with the large $N$ universality of the Durhuus-Olesen large $N$
phase transition in three Euclidean dimensions. However, it has turned out to
be too difficult to determine the large $N$ critical exponents from the data.
If we assume they are $1/2,3/4$ we get consistency, but this is a somewhat
weaker test then getting the exponents from the data at the expected
values, with small errors. This type of numerical problem is common.

In practice one does not need to use the definition of $a(b)$ from the
string tension as presented in the test. In $d=3$ dimensions, as $b\to\infty$, 
the asymptotic perturbative of $a(b)$ is very simple:
\be
a(b)\sim c/b~{\rm for~}b\to\infty
\ee
There is a procedure to replace the constant $c$ 
above by a function $c(b)$
with $c(\infty)=c$. The function $c(b)$ is 
easy to extract from the Monte Carlo simulation, 
as it is directly related to
the action. 
The large $b$ asymptotic regime is entered for values 
of $b$ that are much
smaller than when $c(b)$ is replaced by the constant 
$c$. This technique is called 
``tadpole improvement'' or ``mean field improvement''. To test for
the existence of a continuum limit, it is sufficient to use $a(b)$ as
defined above, and one does not need to also find $\Sigma(b)$. One can find
in the literature various determinations of $\Sigma(b)$ which can be 
extrapolated to the range one uses and convert any continuum numbers that
admit a dimensional interpretation into $MeV$ units. There is no meaning to
the actual number, as the string tension value we can assign a dimensional value
to is the four dimensional one at $N=3$.

In $d=4$, so far we have only a partial test: We know that smeared loops 
have a gap opening transition at a physical length at infinite $N$, but
we have not tested for the critical exponents 1/2 and 3/4. Our guess is that 
consistent results similar to those in tree dimensions would be obtained.

\section{The bigger picture}

On the lattice, one can define another ``string tension'', which is
loop size dependent, by using a Creutz ratio:
\be
\Sigma_{Creutz}(L,b)=-\log\left ( \frac{\langle tr W(L,L+1) \rangle 
\langle tr W(L+1,L) \rangle}{\langle tr W(L,L) \rangle 
\langle tr W(L+1,L+1) \rangle }\right )
\label{creutz}
\ee
where $W_f(L_1,L_2)$ is the Wilson loop matrix for an $L_1\times L_2$. 

$\Sigma_{Creutz} (L,b^{-1} )$ can 
be expanded for $b^{-1}\to 0$ (the regular Feynman expansion) 
and for $b\to 0$ (the so called ``lattice 
strong coupling expansion''. 
The regimes
of validity of these two expansions are disjoint; in between there is a crossover
regime and we can bridge it only by numerical calculation. 

There are extra
complications around $b=0$.  Rectangular loops 
of the type usually used have a ``roughening'' non analyticity in $b$ at
a point $b_{\rm rough}$. 
This non-analyticity is a lattice artifact. 
It can be avoided by choosing loops at generic angles with lattice planes.
Then, the definition of $\Sigma_{Creutz}$ needs to extended. All this
will increase the complexity of the strong coupling expansion. In the end, 
only a physical crossover separating the ranges of 
the weak and strong coupling expansion
remains. We have no non-numerical calculational method to bridge it.
To get the continuum string tension in units
of the perturbative 
scale $\Lambda_{SU(N)}$ we need to take the continuum limit, which is, as we have seen, 
a correlated limit
in which $b^{-1}\to 0$ and the overall lattice scale of the loop goes to infinity. This correlated 
limit preserves the crossover.

The idea we are pursuing is to improve the above scheme in two respects.
First, since we wish to set up a calculation in continuum we forget about the
lattice. Instead of thinking about $\Sigma_{Creutz}$ we consider some other 
observable, for definiteness the extremal eigenvalue $\theta_M$ of a Wilson
loop of size $l$. 

For this to make sense, we need to be able to define
$\theta_M$ in renormalized continuum field theory. We hope that this can be done
by first constructing a renormalized polynomial in $z$ corresponding to 
$\langle \det(z - W_f)\rangle$ (for example by using smearing) 
and taking the roots of it to define $\theta_M$.
While we have some idea how a calculation for small loops might proceed,
for large loops we need something beyond ordinary field theory. Here we assume
that an effective string model will describe $\langle \det(z - W_f)\rangle$. This
model will have a dimensional parameter, the string tension, and will be a good
description for very large loops, with corrections parametrized by an increasing
number of dimensionless parameters
becoming more and more important as the loop shrinks. 

To relate the
string tension to $\Lambda_{SU(N)}$, the dimensional parameter entering the perturbation
theory for small loops, one needs to join the two regimes over the crossover.
Here is the point that the simplification of large $N$ enters: At infinite $N$
the crossover for $z_M$ collapses into a point and we have a
phase transition. We postulate that we know that the transition is universal
and that we know it is in the same universality class at the Durhuus
Olesen transition. This postulate has good reasons to be correct, as I
described. 

Therefore, for $N>>1$, the dependence of $z_M$ on intermediate scales, that 
is scales in the vicinity of the critical scale is known up to a few constants. 
This is the ingredient that was missing in the lattice scenario
described above. It is now possible to imagine calculating
to some order at short, intermediate and long scales and sewing together the
three scale ranges. Requiring smooth matches could produce a number for
the string tension in units of the perturbative scale $\Lambda_{SU(N)}$.

There are many variations possible. $z_M$ is only 
one possible example of a potentially useful variable. $z_M$ 
depends on the dilation of a fixed shaped Wilson loop, 
measured by $l$. As a function of $l$, $z_M$  will trace out
a trajectory from $\theta=0$ at $l=0$  (one could replace this
by a $0<l\Lambda_{SU(N)} \ll 1$) 
to $\theta=\pi(1-1/(2N))$ at $l=\infty$. For small $l$, 
the perturbative scale $\Lambda_{SU(N)}$ enters the
calculations, and for large $l$ the string tension enters.
The two regimes are joined by the crossover. These lectures focused on this
crossover, and this section has sketched the intuitive motivation for 
studying this crossover.

\subsection*{Acknowledgments}
I acknowledge partial support by the DOE under grant
number DE-FG02-01ER41165 at Rutgers University and by the SAS of Rutgers University.
I note with regret that my research has for a long time been 
deliberately obstructed by my high energy colleagues at Rutgers.  
Many thanks are due to my collaborators, 
R. Lohmayer, R. Narayanan, T. Wettig.
I am grateful to A. Schwimmer and L. Stodolsky for illuminating discussion. 
I have greatly benefited from conversations with, 
and communications by J-P Blaizot and M. Nowak. 

I wish the express my gratitude to the organizers of the school for providing
me with the opportunity to present the research program I have been involved
in for the last few years and for their warm hospitality.

\end{document}